\documentclass[preprintnumbers,amsmath,amssymbm,prd]{revtex4}
\usepackage{epsfig}
\usepackage{graphicx}

\begin{document}
\title{Onset of spontaneous scalarization in spinning Gauss-Bonnet black holes}
\author{Shahar Hod}
\affiliation{The Ruppin Academic Center, Emeq Hefer 40250, Israel}
\affiliation{ }
\affiliation{The Hadassah Institute, Jerusalem 91010, Israel}
\date{\today}

\begin{abstract}
\ \ \ It has recently been proved numerically that spinning black holes
in Einstein-scalar theories which are characterized
by a non-minimal negative coupling of the scalar field to the
Gauss-Bonnet invariant of the curved spacetime may develop
exponentially growing instabilities. Intriguingly,
it has been demonstrated that this tachyonic instability,
which marks the onset of the spontaneous scalarization phenomenon
in the Einstein-Gauss-Bonnet-scalar theory, characterizes spinning
black holes whose dimensionless angular momentum parameter ${\bar a}\equiv a/M$
is larger than some critical value ${\bar a}_{\text{crit}}\simeq0.505$.
In the present paper we prove, using {\it analytical} techniques,
that the critical rotation parameter
which marks the boundary between bald Kerr black holes and
hairy (scalarized) spinning black holes in the Einstein-Gauss-Bonnet-scalar theory is
given by the exact dimensionless relation ${\bar a}_{\text{crit}}={1\over 2}$.
\end{abstract}
\bigskip
\maketitle

\section{Introduction}

Various no-hair theorems have established the physically interesting
fact that, within the framework of classical general relativity,
black holes with spatially regular horizons cannot support static
scalar hairy configurations which are minimally coupled to gravity
\cite{Bek1,Sot2,Her1,Sot3}. This no-hair property also characterizes
the black-hole solutions of the non-linearly coupled Einstein-scalar
field equations in which the scalar fields are non-minimally coupled
to the Ricci scalar $R$ of the corresponding curved spacetimes
\cite{BekMay,Hod1}.

Intriguingly, it has recently been revealed that spatially regular
scalar fields which are non-minimally coupled to the Gauss-Bonnet
curvature invariant ${\cal G}$ can be supported in static
\cite{Sot5,Sot1,GB1,GB2} and stationary \cite{ChunHer,SotN}
black-hole spacetimes (see \cite{Hersc1,Hersc2,Hodsc1,Hodsc2} for
the closely related phenomenon of charged black holes that support
spatially regular scalar fields which are non-minimally coupled to
the electromagnetic tensor of the charged spacetime).

In particular, it has been explicitly proved in the physically
important works \cite{Sot5,Sot1,GB1,GB2,ChunHer,SotN} that, within
the extended framework of a Scalar-Tensor-Gauss-Bonnet theory whose
action contains a direct coupling term of the form $f(\phi){\cal G}$
between the scalar field $\phi$ and the Gauss-Bonnet curvature
invariant ${\cal G}$, the canonical Schwarzschild and Kerr
black-hole solutions of the Einstein field equations may become
unstable to linearized perturbations of the non-minimally coupled
scalar fields. This is a tachyonic instability which is
characterized by the presence of an effective {\it negative}
(squared) mass term of the form $-\eta\phi{\cal G}$ in the wave
equation of the linearized non-minimally coupled scalar field, where
the physical parameter $\eta$ controls the strength of the
non-minimal coupling between the scalar field and the Gauss-Bonnet
curvature invariant in the linearized regime [see Eq. (\ref{Eq7})
below].

The exponentially growing tachyonic instabilities of the linearized
non-minimally coupled scalar fields in the Schwarzschild and Kerr
\cite{Notebk} black-hole spacetimes mark the onset of the {\it
spontaneous scalarization} phenomenon in the extended
Einstein-Gauss-Bonnet-scalar theory
\cite{Sot5,Sot1,GB1,GB2,ChunHer,SotN}. In particular, as explicitly
demonstrated in \cite{Sot5,Sot1,GB1,GB2,ChunHer,SotN}, the boundary
between bald (scalarless) black-hole solutions of the field
equations and hairy (scalarized) composed black-hole-scalar-field
configurations in the Scalar-Tensor-Gauss-Bonnet theory is marked by
the presence of Schwarzschild (or Kerr) black holes that support
linearized spatially regular configurations of the non-minimally
coupled scalar fields (see also \cite{Hodsg1,Hodsg2}).

The spontaneous scalarization phenomenon of {\it spinning} black
holes in the Einstein-Gauss-Bonnet-scalar theory with positive
values of the coupling parameter $\eta$ has been studied in the
physically important work \cite{ChunHer}. Most recently, Dima,
Barausse, Franchini, and Sotiriou \cite{SotN} have studied
numerically the (in)stability properties of spinning Kerr black
holes to linearized perturbations of the non-minimally coupled
scalar fields with {\it negative} values of the physical coupling
parameter $\eta$. Intriguingly, it has been revealed in \cite{SotN}
that the non-minimally coupled Einstein-Gauss-Bonnet-scalar theory
with $\eta<0$ is characterized by the presence of a critical
black-hole rotation parameter \cite{Noteaa},
\begin{equation}\label{Eq1}
\Big({{a}\over{M}}\Big)_{\text{crit}}\simeq 0.505\ ,
\end{equation}
which separates stable Kerr black holes from rapidly spinning Kerr
black holes that develop exponentially growing instabilities in
response to linearized perturbations of the non-minimally coupled
scalar fields.

The physically interesting numerical results presented in
\cite{SotN} implies, in particular, that black holes with
sufficiently high spins ($a\geq a_{\text{crit}}\simeq 0.505M$) in the
non-minimally coupled Einstein-Gauss-Bonnet-scalar theory with
negative values of the physical coupling parameter $\eta$ are
expected to be characterized by the intriguing spontaneous
scalarization phenomenon.

The main goal of the present paper is to explore, using analytical
techniques, the onset of the spontaneous scalarization phenomenon
(the onset of the tachyonic instabilities) in {\it spinning}
black-hole spacetimes of the extended Scalar-Tensor-Gauss-Bonnet
theory with negative values of the physical coupling parameter
$\eta$. In particular, below we shall explicitly prove that the
critical rotation parameter $(a/M)_{\text{crit}}\simeq0.505$ that was
first computed numerically in the physically interesting work
\cite{SotN}, and which marks the boundary between bald (scalarless)
Kerr black holes and hairy (scalarized) spinning black-hole
spacetimes of the non-minimally coupled Einstein-Gauss-Bonnet-scalar
theory, can be determined {\it analytically}.

\section{Description of the system}

We shall analyze the onset of the spontaneous scalarization
phenomenon in the composed Einstein-Gauss-Bonnet-scalar theory in
which the scalar field is characterized by a non-minimal negative
coupling to the Gauss-Bonnet curvature invariant of a spinning Kerr
black hole. The curved black-hole spacetime is described by the line
element \cite{ThWe,Chan,Noteun,Notebl}
\begin{eqnarray}\label{Eq2}
ds^2=-{{\Delta}\over{\rho^2}}(dt-a\sin^2\theta
d\phi)^2+{{\rho^2}\over{\Delta}}dr^2+\rho^2
d\theta^2+{{\sin^2\theta}\over{\rho^2}}\big[a
dt-(r^2+a^2)d\phi\big]^2\  ,
\end{eqnarray}
where $M$ and $a$ are respectively the mass and angular momentum per
unit mass of the black hole. The metric functions in (\ref{Eq2}) are
given by the functional expressions $\Delta\equiv r^2-2Mr+a^2$ and
$\rho^2\equiv r^2+a^2\cos^2\theta$. The outer and inner horizon
radii of the Kerr black-hole spacetime are determined by the roots
of the metric function $\Delta$:
\begin{equation}\label{Eq3}
r_{\pm}=M\pm(M^2-a^2)^{1/2}\  .
\end{equation}

The composed Einstein-Gauss-Bonnet-scalar theory is characterized by the action
\cite{GB1,GB2}
\begin{equation}\label{Eq4}
S={1\over2}\int
d^4x\sqrt{-g}\Big[R-{1\over2}\nabla_{\alpha}\phi\nabla^{\alpha}\phi+f(\phi){\cal
G}\Big]\  .
\end{equation}
Here
\begin{equation}\label{Eq5}
{\cal G}\equiv R_{\mu\nu\rho\sigma}R^{\mu\nu\rho\sigma}-4R_{\mu\nu}R^{\mu\nu}+R^2\
\end{equation}
is the Gauss-Bonnet invariant of the curved spacetime.
For a spinning Kerr black hole, the Gauss-Bonnet invariant
is given by the ($r$ and $\theta$ dependent) functional expression \cite{SotN}
\begin{equation}\label{Eq6}
{\cal G}_{\text{Kerr}}(r,\theta)={{48M^2}\over{(r^2+a^2\cos^2\theta)^6}}
\cdot\big({r^6-15a^2r^4\cos^2\theta+15a^4r^2\cos^4\theta-a^6\cos^6\theta}\big)\  .
\end{equation}

The term $f(\phi){\cal G}$ in the action (\ref{Eq4}) controls the
non-minimal coupling between the scalar field and the Gauss-Bonnet
invariant (\ref{Eq6}) of the curved and spinning black-hole
spacetime. In the linearized regime, the scalar coupling function
has the universal quadratic behavior \cite{GB1,GB2,Hersc1}
\begin{equation}\label{Eq7}
f(\phi)={1\over2}\eta\phi^2\  ,
\end{equation}
where the physical parameter $\eta$, which can be positive
\cite{ChunHer} or negative \cite{SotN}, controls the strength of the
non-minimal coupling between the scalar field and the Gauss-Bonnet
curvature invariant \cite{Noteed}. As explicitly shown in
\cite{GB1,GB2,Hersc1}, the simple quadratic form (\ref{Eq7}) of the
scalar coupling function guarantees that the bald (scalarless)
Schwarzschild and Kerr black-hole spacetimes are valid solutions of
the Einstein field equations in the trivial $\phi\to0$ limit.

A variation of the action (\ref{Eq4}), which characterizes the
non-minimally coupled Einstein-Gauss-Bonnet-scalar theory, with
respect to the scalar field $\phi$ yields the differential equation
\cite{Noteee}
\begin{equation}\label{Eq8}
\nabla^\nu\nabla_{\nu}\phi=-f'(\phi){\cal G}\  ,
\end{equation}
which determines the spatial and temporal behavior of the non-minimally coupled scalar
field in the spinning black-hole spacetime.

As shown in \cite{SotN}, a projection of the differential equation
(\ref{Eq8}) onto a basis of the angular-dependent spherical harmonic
functions $Y_{lm}(\theta,\varphi)$ \cite{SphHar,Notelm} yields, in
the linearized regime, the coupled spatio-temporal differential
equations \cite{SotN}
\begin{eqnarray}\label{Eq9}
[(r^2+a^2)^2-a^2\Delta(1-c^m_{ll})]\ddot\psi_{lm}+a^2\Delta(c^m_{l,l+2}
\ddot\psi_{l+2,m}+c^m_{l,l-2}\ddot\psi_{l-2,m})\\ \nonumber
+4iamMr\dot\psi_{lm}-(r^2+a^2)^2\psi''_{lm}-[2iam(r^2+a^2)-2a^2\Delta/r]\psi'_{lm}\\ \nonumber
+\Delta[l(l+1)+2M/r-a^2/r^2+2iam/r]\psi_{lm}
+\Delta\sum_j\langle lm|\mu^2_{\text{eff}}(r^2+a^2\cos^2\theta)|jm\rangle\psi_{jm} & = & 0\  ,
\end{eqnarray}
where
\begin{equation}\label{Eq10}
\psi_{lm}(t,r)\equiv\int r\phi Y_{lm}Y^*_{lm} d\Omega\equiv\langle lm|r\phi|lm\rangle\
\end{equation}
and \cite{SotN}
\begin{equation}\label{Eq11}
c^m_{jl}\equiv\langle lm|\cos^2\theta|jm\rangle\  .
\end{equation}
Interestingly, the differential equation (\ref{Eq9}) is
characterized by the presence of an effective mass term
$\mu^2_{\text{eff}}$ which is a direct consequence of the
non-minimal coupling between the scalar field and the Gauss-Bonnet
curvature invariant (\ref{Eq6}). This spatially-dependent effective
mass term is given by the mathematical expression \cite{SotN}
\begin{equation}\label{Eq12}
\mu^2_{\text{eff}}=-\eta{\cal G}\  .
\end{equation}

In the next section we shall study analytically the onset of the
spontaneous scalarization phenomenon in the composed
Einstein-Gauss-Bonnet-scalar theory (\ref{Eq4}) with negative values
\cite{SotN} of the non-minimal coupling parameter $\eta$. As
discussed in \cite{Sot5,Sot1,GB1,GB2,ChunHer,SotN}, this intriguing
physical phenomenon is related to the presence of an effective
negative (squared) mass term in the wave equation of the
non-minimally coupled scalar field which, in the vicinity of the
black-hole horizon, acts as an effective binding potential.

\section{Onset of the spontaneous scalarization phenomenon in spinning Gauss-Bonnet black-hole spacetimes}

In the present section we shall determine, using {\it analytical}
techniques, the critical value $(a/M)_{\text{crit}}$ of the
dimensionless black-hole rotation parameter which marks the onset of
tachyonic instabilities in the non-minimally coupled
Einstein-Gauss-Bonnet-scalar theory (\ref{Eq4}) with negative values
of the physical coupling parameter $\eta$. As explicitly
demonstrated in \cite{Sot5,Sot1,GB1,GB2,ChunHer,SotN}, these
tachyonic instabilities are closely related to the intriguing
phenomenon of black-hole spontaneous scalarization. In particular,
the critical rotation parameter $(a/M)_{\text{crit}}$ marks the
boundary between bald Kerr black holes and hairy (scalarized)
spinning black holes in the Einstein-Gauss-Bonnet-scalar theory
(\ref{Eq4}) with negative values of the coupling parameter $\eta$
\cite{Noteexp}.

As nicely demonstrated numerically in \cite{SotN}, for a given value
of the Kerr black-hole rotation parameter $a$, the onset of the
tachyonic instabilities in the composed
black-hole-nonminimally-coupled-linearized-scalar-field system is
marked by the presence of {\it marginally-stable} scalar
modes with
\begin{equation}\label{Eq13}
m^*=0\
\end{equation}
which are characterized by an infinitely long instability timescale.
In particular, at the onset of the
tachyonic instability, the mixed term $\Delta\sum_j\langle
lm^*|\mu^2_{\text{eff}}(r^2+a^2\cos^2\theta)|jm^*\rangle\psi_{jm^*}$ in Eq. (\ref{Eq9}) can be
replaced by a single term of the form
\begin{equation}\label{Eq14}
\Delta\langle l_1m^*=0|\mu^2_{\text{eff}}(r^2+a^2\cos^2\theta)|l_2m^*=0\rangle\psi_{l_1m^*=0}\
\end{equation}
at asymptotically late times.

Interestingly, it has been revealed in
\cite{Sot5,Sot1,GB1,GB2,ChunHer,SotN} that the physically intriguing
phenomenon of black-hole spontaneous scalarization is related to the
presence of an effective binding potential well [an effective {\it
negative} (squared) mass term, see Eqs. (\ref{Eq12}) and
(\ref{Eq14})] in the vicinity of the black-hole outer horizon whose
two turning points $\{r_{\text{in}},r_{\text{out}}\}$ are
characterized by the relation $r_{\text{out}}\geq
r_{\text{in}}=r_+$. In particular, the marginally-stable critical
black hole (with $a=a_{\text{crit}}$), which marks the boundary
between bald Kerr black holes and hairy (scalarized) spinning black
holes in the Einstein-Gauss-Bonnet-scalar theory (\ref{Eq4}) with
negative values of the coupling parameter $\eta$, is characterized
by the presence of a {\it degenerate} binding potential well whose
two turning points merge at the outer horizon of the black hole
\cite{Noteetainf}:
\begin{equation}\label{Eq15}
\langle l_1m^*=0|\mu^2_{\text{eff}}(r^2+a^2\cos^2\theta)
|l_2m^*=0\rangle_{r=r_{\text{in}}=r_{\text{out}}=r_+(a_{\text{crit}})}=0\
\ \ \ \text{for}\ \ \ \ a=a_{\text{crit}}\  .
\end{equation}

Taking cognizance of Eqs. (\ref{Eq3}), (\ref{Eq6}), (\ref{Eq10}),
(\ref{Eq12}), (\ref{Eq13}), and (\ref{Eq15}), one finds that the
critical rotation parameter $a_{\text{crit}}$ of the non-minimally
coupled Einstein-Gauss-Bonnet-scalar theory (\ref{Eq4}) with
$\eta<0$ can be determined by the (rather cumbersome) integral relation
\begin{equation}\label{Eq16}
\int_{0}^{\pi}{{r^6_+-15a^2r^4_+\cos^2\theta+15a^4r^2_+\cos^4\theta-a^6\cos^6\theta}
\over{(r^2_++a^2\cos^2\theta)^5}}\cdot Y_{l_1m^*=0}(\cos\theta)Y_{l_2m^*=0}(\cos\theta)
\sin\theta d\theta=0\ \ \ \ \text{for}\ \ \ \ a=a_{\text{crit}}\  .
\end{equation}
Interestingly, and most importantly for our analysis, the resonance equation (\ref{Eq16}) 
can be solved {\it analytically}. 

To see this, it proves useful to
define the dimensionless variables
\begin{equation}\label{Eq17}
{\hat a}\equiv {{a}_{\text{crit}}\over{r_+}}\
\end{equation}
and
\begin{equation}\label{Eq18}
x\equiv {\hat a}\cdot\cos\theta\  ,
\end{equation}
in terms of which the characteristic integral equation (\ref{Eq16})
can be written in the form
\begin{equation}\label{Eq19}
\int_{0}^{\hat a}{{1-15x^2+15x^4-x^6}\over{(1+x^2)^5}}\cdot Y_{l_1m^*=0}(x/{\hat a})Y_{l_2m^*=0}(x/{\hat a}) dx=0\  .
\end{equation}
The integral (\ref{Eq19}) can be evaluated analytically. To illustrate this, we shall now present some examples:
\newline
(1) For $(l_1,l_2)=(0,0)$ one obtains from the integral (\ref{Eq19}) 
the resonance equation
\begin{equation}\label{Eq20}
{\hat a}^4-8{\hat a}^2+3=0\  ,
\end{equation}
whose solution is given by the compact dimensionless expression 
\begin{equation}\label{Eq21}
{\hat a}=\sqrt{4-\sqrt{13}}\  .
\end{equation}
Taking cognizance of Eqs. (\ref{Eq3}), (\ref{Eq17}), and (\ref{Eq21}), one finds
\begin{equation}\label{Eq22}
\Big({{a}\over{M}}\Big)_{\text{crit}}=\sqrt{{{11+\sqrt{13}}\over{18}}}\simeq0.9008\ \ \ \ {\text for}\ \ \ \ (l_1,l_2)=(0,0)\  .
\end{equation}
It is worth emphasizing that the analytically derived value (\ref{Eq22}) agrees remarkably 
well with the numerical results presented in \cite{SotNv1} (see, in particular, Fig. 2 of \cite{SotNv1}). 
\newline
(2) For $(l_1,l_2)=(1,1)$ one obtains from the integral (\ref{Eq19}) 
the resonance equation
\begin{equation}\label{Eq23}
3{\hat a}^4-8{\hat a}^2+1=0\  ,
\end{equation}
whose solution is given by the compact dimensionless expression 
\begin{equation}\label{Eq24}
{\hat a}=\sqrt{{{4-\sqrt{13}}\over{3}}}\  .
\end{equation}
Taking cognizance of Eqs. (\ref{Eq3}), (\ref{Eq17}), and (\ref{Eq24}), one finds
\begin{equation}\label{Eq25}
\Big({{a}\over{M}}\Big)_{\text{crit}}=\sqrt{{{11-\sqrt{13}}\over{18}}}\simeq0.6409\ \ \ \ {\text for}\ \ \ \ (l_1,l_2)=(1,1)\  .
\end{equation}
It is worth emphasizing again that the analytically derived value (\ref{Eq25}) agrees remarkably 
well with the numerical results presented in \cite{SotNv1} (see, in particular, Fig. 1 of \cite{SotNv1}). 

Interestingly, in the asymptotic $l_1=l_2\to\infty$ limit the 
spherical harmonic function $Y_{lm=0}(\cos\theta)$ 
approaches a delta-function which is peaked around the poles $\theta=0,\pi$ 
(or, equivalently, around $x=\pm{\hat a}$) \cite{Notespll}. 
In this large-$l$ limit the integral equation (\ref{Eq19}) yields the remarkably simple resonance relation
\begin{equation}\label{Eq29}
1-15{\hat a}^2+15{\hat a}^4-{\hat a}^6=0\  ,
\end{equation}
whose solution is given by
\begin{equation}\label{Eq30}
{\hat a}=2-\sqrt{3}\  .
\end{equation}
Taking cognizance of Eqs. (\ref{Eq3}), (\ref{Eq17}), and (\ref{Eq30}), one finds the simple dimensionless expression
\begin{equation}\label{Eq31}
\Big({{a}\over{M}}\Big)_{\text{crit}}={1\over2}\ \ \ \ {\text for}\ \ \ \ l^*_1=l^*_2\to\infty\
\end{equation}
for the critical black-hole rotation parameter. 

Remarkably, the analytically derived critical black-hole spin
parameter (\ref{Eq31}) agrees extremely well with the critical value
(\ref{Eq1}) that was first computed numerically in the physically
important work \cite{SotN}. As nicely shown numerically in
\cite{SotN}, the critical black-hole rotation parameter
$a_{\text{crit}}$ marks the onset of tachyonic instabilities (the
onset of the spontaneous scalarization phenomenon) in the
non-minimally coupled Einstein-Gauss-Bonnet-scalar theory
(\ref{Eq4}). In particular, slowly spinning Kerr black holes with
rotation parameters in the regime $a<a_{\text{crit}}$ do not develop
tachyonic instabilities and cannot carry spatially regular scalar
hairy configurations in the field theory (\ref{Eq4}) with negative
values of the physical coupling parameter $\eta$.

\section{Summary}

It has recently been proved \cite{Sot5,Sot1,GB1,GB2,ChunHer,SotN}
that black holes in Einstein-Gauss-Bonnet-scalar theories in which
the scalar field is non-minimally coupled to the Gauss-Bonnet
invariant of the curved spacetime can support spatially regular
bound-state configurations of the scalar field.

In particular, the recently published important work \cite{SotN} has
revealed the physically intriguing fact that in
Einstein-Gauss-Bonnet-scalar theories with negative values of the
physical coupling parameter $\eta$, there exists a critical
black-hole rotation parameter [see Eq. (\ref{Eq1})],
\begin{equation}\label{Eq32}
{\bar
a}_{\text{crit}}\equiv\Big({{a}\over{M}}\Big)_{\text{crit}}\simeq0.505\  ,
\end{equation}
which separates stable Kerr black holes from rapidly spinning black
holes that develop exponentially growing tachyonic instabilities in
response to linearized perturbations of the non-minimally coupled
scalar fields.

Motivated by the intriguing numerical observation (\ref{Eq32})
presented in \cite{SotN}, in the present paper we have used
analytical techniques in order to study the onset of the spontaneous
scalarization phenomenon of {\it spinning} black holes in the
non-minimally coupled Einstein-Gauss-Bonnet-scalar theory
(\ref{Eq4}) with negative values of the coupling parameter $\eta$.

In particular, we have proved that the critical black-hole rotation
parameter ${\bar a}_{\text{crit}}$, which marks the boundary between
bald Kerr black holes and hairy (scalarized) spinning black holes in
the Einstein-Gauss-Bonnet-scalar theory, is given by the simple
dimensionless relation [see Eq. (\ref{Eq31})]
\begin{equation}\label{Eq33}
{\bar a}_{\text{crit}}={1\over 2}\ .
\end{equation}
It is interesting to note that the {\it analytically} derived
critical black-hole rotation parameter (\ref{Eq33}) agrees
remarkably well with the corresponding {\it numerically} computed
critical rotation parameter (\ref{Eq32}) of \cite{SotN}.

Finally, it is worth emphasizing again that the physical
significance of the critical black-hole rotation parameter
(\ref{Eq33}) stems from the fact that it marks the boundary between
bald (scalarless) Kerr black holes and hairy (scalarized) black-hole
solutions of the non-minimally coupled Einstein-Gauss-Bonnet-scalar
theory with negative values of the physical coupling parameter
$\eta$ \cite{Notesh}.

\newpage

\bigskip
\noindent
{\bf ACKNOWLEDGMENTS}
\bigskip

This research is supported by the Carmel Science Foundation. I would
like to thank Alexandru Dima, Enrico Barausse, Nicola Franchini, and Thomas Sotiriou for 
an important correspondence. I would also like to thank 
Yael Oren, Arbel M. Ongo, Ayelet B. Lata, and Alona B. Tea for helpful discussions.



\begin{thebibliography}{99}

\bibitem{Bek1} J. D. Bekenstein, Phys. Rev. D {\bf 5}, 1239 (1972).

\bibitem{Sot2} T. P. Sotiriou, Class. Quant. Grav. {\bf 32}, 214002 (2015).

\bibitem{Her1} C. A. R. Herdeiro and E. Radu, Int. J. Mod. Phys. D {\bf 24}, 1542014 (2015).

\bibitem{Sot3} T. P. Sotiriou and V. Faraoni, Phys. Rev. Lett. {\bf 108}, 081103 (2012).

\bibitem{BekMay} A. E. Mayo and J. D. Bekenstein and, Phys. Rev. D {\bf 54}, 5059 (1996).

\bibitem{Hod1} S. Hod, Phys. Lett. B {\bf 771}, 521 (2017) [arXiv:1911.08371];
S. Hod, Phys. Rev. D {\bf 96}, 124037 (2017) [arXiv:2002.05903].

\bibitem{Sot5} T. P. Sotiriou and S.-Y. Zhou, Phys. Rev. Lett. {\bf 112}, 251102
(2014); T. P. Sotiriou and S.-Y. Zhou, Phys. Rev. D {\bf 90}, 124063
(2014).

\bibitem{Sot1} T. P. Sotiriou, Lect. Notes Phys. {\bf 892}, 3 (2015) [arXiv:1404.2955].

\bibitem{GB1} D. D. Doneva and S. S. Yazadjiev, Phys. Rev. Lett. {\bf 120}, 131103 (2018).

\bibitem{GB2} H. O. Silva, J. Sakstein, L. Gualtieri, T. P. Sotiriou, and E. Berti,
Phys. Rev. Lett. {\bf 120}, 131104 (2018).

\bibitem{ChunHer} P. V. P. Cunha, C. A. R. Herdeiro, and E. Radu, Phys. Rev. Lett. {\bf 123}, 011101 (2019).

\bibitem{SotN} A. Dima, E. Barausse, N. Franchini, and T. P. Sotiriou, arXiv:2006.03095v2 .

\bibitem{Hersc1} C. A. R. Herdeiro, E. Radu, N. Sanchis-Gual, and J. A. Font, Phys.
Rev. Lett. {\bf 121}, 101102 (2018).

\bibitem{Hersc2} P. G. S. Fernandes, C. A. R. Herdeiro, A. M. Pombo, E. Radu, and N. Sanchis-Gual,
Class. Quant. Grav. {\bf 36}, 134002 (2019) [arXiv:1902.05079].

\bibitem{Hodsc1} S. Hod, Phys. Lett. B {\bf 798}, 135025 (2019) [arXiv:2002.01948].

\bibitem{Hodsc2} S. Hod, Phys. Rev. D {\bf 101}, 104025 (2020) [arXiv:2005.10268].

\bibitem{Notebk} To the best of our knowledge, most studies in the physics literature
of the tachyonic instabilities of black holes to scalar
perturbations in extended Scalar-Tensor-Gauss-Bonnet theories have
focused on the case of spherically symmetric spacetimes. The
physically interesting case of non-spherically symmetric spinning
black holes in extended Scalar-Tensor-Gauss-Bonnet theories has been
studied in \cite{ChunHer,SotN}.

\bibitem{Hodsg1} S. Hod, Phys. Rev. D {\bf 100}, 064039 (2019) [arXiv:1912.07630].

\bibitem{Hodsg2} S. Hod, The Euro. Phys. Jour. C {\bf 79}, 966 (2019).

\bibitem{Noteaa} Here $M$ is the black-hole mass and
$a\equiv J/M$ is the angular momentum per unit mass of the black
hole. We shall henceforth assume $a>0$ without loss of generality.

\bibitem{ThWe} C. W. Misner, K. S. Thorne, and J. A. Wheeler, {\it Gravitation},
(W. H. Freeman, San Francisco, 1973).

\bibitem{Chan} S. Chandrasekhar, {\it The Mathematical Theory of Black
Holes}, (Oxford University Press, New York, 1983).

\bibitem{Noteun} We shall use natural units in which $8\pi G=c=\hbar=1$.

\bibitem{Notebl} Here $(t,r,\theta,\phi)$ are the Boyer-Lindquist spacetime coordinates.

\bibitem{Noteed} Note that $\eta$, the physical coupling parameter of the theory,
has the dimensions of length$^2$.

\bibitem{Noteee} Here $f'\equiv df/d\phi$.

\bibitem{SphHar} D. A. Varshalovich, A. N. Moskalev, and V. K. Khersonskii,
{\it Quantum Theory of Angular Momentum} (World Scientific, 1988).

\bibitem{Notelm} The angular parameters $\{l,m\}$ are characterized by the relation $l\geq|m|$.

\bibitem{Noteexp} Note that spinning black holes with non-minimally coupled scalar hair
and a negative coupling parameter $\eta$ are expected to be characterized by the
relation $a\geq a_{\text{crit}}$ \cite{SotN}.

\bibitem{Noteetainf} It is worth noting that, as explicitly demonstrated numerically in \cite{SotN},
the critical boundary (with $a=a_{\text{crit}}$) between stable and
unstable Kerr black holes in the composed theory (\ref{Eq4}) with
negative values of the physical coupling parameter $\eta$ is
characterized by the relation $\eta\to -\infty$. Thus, in the $a\to
a_{\text{crit}}$ $(\eta\to -\infty)$ limit the last term in Eq.
(\ref{Eq9}), which is proportional to the coupling parameter $\eta$
[see Eqs. (\ref{Eq9}) and (\ref{Eq12})], dominates the spatial
behavior of the effective potential well [the effective (squared)
mass term] in the vicinity of the black-hole horizon.

\bibitem{SotNv1} A. Dima, E. Barausse, N. Franchini, and T. P. Sotiriou, arXiv:2006.03095v1 .

\bibitem{Notespll} In particular, one finds the functional behavior 
$Y_{l\to\infty m=0}(\theta=0)=\sqrt{l/2\pi}\to\infty$ for 
the spherical harmonic functions in the asymptotic $l\to\infty$ limit. 

\bibitem{Notesh} In particular, as emphasized in \cite{SotN}, spinning Kerr black holes
with rotation parameters in the regime $a<a_{\text{crit}}$ do not
develop tachyonic instabilities to perturbations of the
non-minimally coupled scalar fields with negative values of the
physical coupling parameter $\eta$. Thus, these black holes (with
$a<a_{\text{crit}}$) are not expected to support spatially regular
hairy configurations of the non-minimally coupled scalar fields.

\end{thebibliography}
\end{document}